\documentclass[aps,prl,twocolumn,groupedaddress,showpacs]{revtex4}
\usepackage{graphicx}
\begin{document}
\title{Corrections to Scaling in the Droplet Picture of Spin Glasses }
\author{M.A.Moore}
\affiliation{Department of Physics and Astronomy, University of 
Manchester,
Manchester, M13 9PL, United Kingdom}
\date{\today}
\begin{abstract}
The energy of a droplet of linear extent $l$ in the ``droplet theory''
of spin glasses goes as $l^{\theta}$ for large $l$. It is argued that this 
formula needs to be modified by the addition of a scaling correction 
$l^{-\omega}$ in order to accurately describe droplet energies at the length
scales currently probed in numerical simulations. With this simple modification
all equilibrium numerical data on Ising spin glasses in two, three and four
dimensions becomes compatible with the droplet model.  
\end{abstract}
\maketitle
The controversy as to the nature of the ordered state of spin glasses still
remains to be resolved. There are two theories: the ``droplet theory'' \cite{
McM,BM,FH} and the replica symmetry breaking (RSB) theory of Parisi
\cite{Parisi,sgb}. Many numerical studies have been done in an attempt
to resolve the controversy with results which are often uncomfortable for
proponents of both theories. The most natural summary of the current 
numerical situation is the TNT (\textit{trivial, non-trivial}) picture of
 Refs. \cite{KM,PY}. On this picture,
which we shall explain in more detail below, the Parisi spin overlap function 
$P(q,L)$ for a system of linear dimension $L$ seems to behave as expected
from the RSB theory i.e. in a \textit{non-trivial} fashion, while the
behavior of the link overlap  seems to be more in accordance with droplet model
ideas as its variance seems to be tending to zero  as $L$ 
increases to infinity (which is regarded as \textit{trivial}). In this paper
we shall show that a simple and natural correction to scaling term added to
the usual droplet energy expression can explain the origin of the TNT picture.

In the RSB theory there are low-energy excitations in which a 
finite fraction of the 
spins in the system are reversed but which only cost a finite amount of energy
in the thermodynamic limit. The surface of these excitations is space-filling 
so that the fractal dimension of their surface $d_s$ is the same as the
space dimension $d$ \cite{Marinari}. In the droplet theory, 
the lowest energy excitations of linear spatial extent $l$ typically costs
of order $l^{\theta}$, where $\theta$ is a positive exponent, of order 0.20
in three dimensions \cite{3D} and 0.70 in four dimensions \cite{4D} 
according to studies of the energy to create a domain wall right
across the system. Thus in
the thermodynamic limit the excitations which flip a finite fraction of the
spins cost an infinite amount of energy and also the surface of these 
excitations is not space filling, as $d_s<d$. 

The Hamiltonian which is usually studied in numerical work is the Ising
spin glass model:
\begin{equation}
{\mathcal H}=-\sum_{<i,j>}J_{ij}S_iS_j,
\end{equation}
where the sites $i$ lie on the sites of a cubic lattice with $N=L^d$ sites,
$S_i=\pm1$, the $J_{ij}$ have a Gaussian distribution of zero mean and
unit variance and couple nearest-neighbor sites on the lattice.
Typically for equilibrium studies at low temperatures
(done at temperatures $T$ much less than the critical temperature $T_c$
in order to diminish the complications from critical point effects
\cite{Drossel}), $L\leq14$ for 3D and $L\leq5$ for 4D. By studying two real 
replicas of the system, with spins $S_i$
and $S^{'}_i$ one can derive the Parisi spin overlap function $P(q,L)$, where 
\begin{equation}
q=\frac{1}{N}\sum_{i}S_iS_i^{'}
\end{equation}
and study the properties of link overlaps
\begin{equation}
q_l=\frac{1}{N_b}\sum_{<ij>}S_iS_jS^{'}_iS^{'}_j,
\end{equation}
where $N_b$ is the number of mearest-neighbor bonds on the lattice. On the
droplet picture $P(q,L)$ should tend as $L\rightarrow\infty$ to a delta 
function at $q=q_{EA}$, and $P(q,L)\sim T/L^{\theta^{'}}$ for
all $q\neq q_{EA}$ where the expectation is that $\theta^{'}=\theta$. In fact 
it is found that $\theta^{'}\approx0$ and that  $P(q,L)$ looks very similar to
what is predicted from the mean-field RSB picture \cite{Marinari} in that
$P(q,L)\neq 0$ for all $q\neq q_{EA}$
It is this result which advocates of the droplet model have found hardest to
understand.

Studies of link overlaps are more in
accordance with droplet model expectations. Their variance is predicted
on the droplet model to decrease with $L$ as $T/L^{\mu_l}$ with
$\mu_l=2(d-d_s)+\theta^{'}$ \cite{Drossel2}. The most recent investigations
of link overlaps
have been performed by adding a bulk perturbation $-\epsilon q_l$ to 
${\mathcal H}$ and 
studying its effect in the ground state (see below).
Using this method Palassini and Young
\cite{PY} report that in 3D $d_s= 2.58\pm0.02$ and
$\theta^{'}=0.02\pm0.03$
while in 4D $d_s=3.77\pm0.05$ and $\theta^{'}=0.03\pm0.05$. The droplet
model expectation is again that $\theta^{'}=\theta$, which is clearly not 
satisfied
at the $L$ values which can be currently studied. On the RSB picture
the variance of
$q_l$ should be non-zero for large L, which is not in accord with the results
of \cite{PY}. (However, Marinari and Parisi \cite{MP} did obtain a finite
variance of the link overlap in the large $L$ limit by extrapolating it in an
ad hoc way as $1/L$).

Thus if one accepts the TNT version of the numerical results but believes in 
the droplet model, the task is to understand why $\theta^{'}\ne\theta$
at least at the small values of $L$ which can be currently reached in
equilibrium numerical studies. We will assume that the 
droplet picture is essentially correct and explain why corrections
to scaling result in $\theta{'}$ appearing to be
close to zero for small $L$ values. 

It is important to realise what the low-energy excitations look like in spin
glasses. Consider a domain wall crossing a system  $M\times M\times L$. The
energy cost of this domain on the droplet picture will be of order
$M^{\theta}$ and the domain wall will be fractal with dimension $d_s$ and 
with an area of order $M^{d_s}$. In three or more dimensions the 
interface may have holes throught it. 
Because it is fractal the extent of this wandering
is of order $M$. The wandering of the interface by an amount of order $M$ 
affects the determination of $\theta$, see Ref. \cite{Carter},
where it was found that the best results were obtained when $L\gg M$ as then
the interface is not affected by its interactions with the ends of the system.
Suppose now we have two domain walls across the system. 
Then if their separation 
is large compared to $M$ they will be unaffected by the presence of the other.
However, if they are closer together then they interfere with each other and
their overall energy will be greater than if the other one were absent. In
other words, domain walls effectively repel each other. If they have a 
separation of order $l$, then the repulsive energy between them would
be expected to vary with their separation as a power law,
$l^{-\omega^{'}}$. No investigations of $\omega^{'}$ seem to exist in
the literature.
 
Consider now a droplet of linear extent $l$. It too will have a fractal
surface described by $d_s$. Pictures of large droplets have now appeared 
\cite{KM}, and a systematic investigation of them is in 
\cite{HKM}. Their fractal nature ensures that in three or more
dimensions that they have holes through them,
giving them a sponge-like appearance. However, because the surface of the
droplet may wander by a distance of order $l$ the energy of the droplet $E$
will be modified by its wandering and ``collisions'' with itself to a form 
which we suppose by analogy with the above is 
\begin{equation}
E=Al^{\theta}+Bl^{-\omega}.
\label{basic}
\end{equation}
The term $Bl^{-\omega}$ is a scaling correction to the form of the droplet
energy at large $l$ and $A$ and $B$ are positive constants.
We shall argue that with this scaling correction it is possible to understand
the TNT description of the numerical scene. The exponent $\omega$ is only 
well-defined as a correction to scaling exponent when $l\rightarrow\infty$, but
we shall assume that Eq. (\ref{basic}) has utility outside this limit.
We do not know whether $\omega^{'}=
\omega$, but fortunately there already exists numerical data on the value of
$\omega$ in 3D from the work of Lamarcq et al. \cite{LQ}.

These authors first computed the ground state for both $N=6^3$ and $N=10^3$.
They then chose an arbitrary reference spin and flipped it along
with a cluster containing $v-1$ other spins connected to it. They
next minimized the energy of this cluster by exchange Monte Carlo, but with 
the constraint that the reference spin is held fixed and the cluster was 
always connected and of size $v$. They found the largest extension (mean 
end-to-end distance) $l$ of the 
cluster and found that for $v\leq 33$ its energy $E(l)$ varied as
$l^{-\omega}$, where the exponent $\omega$ was $0.13 \pm0.02$. For larger
$v$ values they found that $E(l)$ was increasing rather than decreasing with
$l$ but they were unsure whether this might not be an artifact of insufficient
numbers of Monte Carlo steps. Note that according to Eq. (\ref{basic}), 
when both
$\theta$ and $\omega$ are small $E(l)$ will have a shallow minimum. The 
clusters generated by their procedure were not compact (that is, $v\sim l^d$),
but on the droplet picture, which is a scaling picture associated with the
zero-temperature fixed point \cite{McM,BM} it is only required that large
clusters be compact.

Eq. (\ref{basic}) is more illuminating when expressed in variables associated
with its minimum at $l=l_0$: 
\begin{equation}
E(l)=C\left(\frac{(l/l_0)^{\theta}}{\theta}+\frac{(l_0/l)^{\omega}}
{\omega}\right),
\end{equation}
where $C$ is a positive constant.
The position of the minimum is essentially unknown but there is a hint 
from the upturn in $E(l)$ seen in Ref. \cite{LQ} that $l_0$ might lie
between 7 and 10 lattice spacings in 3D. Then because the values of both
$\theta$ and $\omega$ are small there is only a 
weak dependence of $E(l)$ on $l$ in the region accessible to
numerical studies (say 14 lattice spacings). 
The minimum is very shallow so that the apparent value of 
$\theta$, which is $\theta^{'}$, will be close to zero. For example,
the ratio $E(14)/E(7)=1.006$ if $l_0$ is 7,
whereas in the absence of the correction to 
scaling term this ratio is 1.15.  In 4D the 
values of $\omega$ and $l_0$ are undetermined at present, but provided $l_0$
happens to lie in the region open to numerically studies, then again 
$\theta^{'}$ would be small.

While it is inevitable that exponents in 4D will be difficult to determine
in numerical studies this is not so for 2D. There seems to be  
indirect evidence which supports an $E(l)$ as in Eq. (\ref{basic}) also
for 2D systems. In two
dimensions $\theta$ is very accurately determined as systems of $480^2$
can be studied \cite{HY}. The
exponent associated with a single domain wall, that is $\theta$, is $-0.282(2)
$ \cite{Carter,HY}. However, studies where, say, $\theta$ is 
determined from
the effects of thermally excited droplets (such as Monte Carlo simulations of
the spin-glass susceptibility)  yield an apparent $\theta^{'}$ close
to $-0.47$ \cite{KA}. This discrepancy has long been a puzzle, and has 
prompted suggestions that perhaps different exponents describe
domain wall energies and droplet energies \cite{KA}. However, Eq. (\ref{basic})
with $\omega\approx0.47$ would seem to offer another way of resolving the
descrepancy. For small values of $l$, droplets would apparently have 
energies decreasing as $l^{-\omega}$ as their energies are dominated
by the correction to scaling term, but at large values of $l$ the decrease
with $l$ will be slower and be as $l^{-|\theta|}$, as expected from 
the conventional droplet approach. 

This crossover between the large and small $l$ dependencies
can be seen in the work of 
Middleton \cite{M}.  He studied
link overlaps as in \cite{PY} and \cite{MP}, by comparing
the unperturbed $J_{ij}$ ground state, $\{S_i^0\}$, with the
$\epsilon$-perturbed state, where 
$J_{ij}\rightarrow J_{ij}-\epsilon S_i^0S_j^0/N_b$. The
fraction of link values $S_iS_j$ on nearest-neighbor
bonds $<ij>$ which are changed by the perturbation is $(1-q_l)$.
On the droplet model the sample average of this,
$\overline{1-q_l}$, is predicted to decrease as $\epsilon/L^{\mu_l}$. Using the
values $\theta=-0.28$ and $d_s=1.27$, $\mu_l\approx1.18$, but 
with $\theta^{'}$ at its small $l$ value of
$-0.47$ the expected value of $\mu_l\approx0.99$.  Middleton studied
the effective value of $\mu_l$ at length scale L using the definition
\begin{equation}
\mu_l^{\rm eff}=-\frac{d[{\rm ln}(\overline{1-q_l})]}{d\, {\rm ln}(L)}.
\end{equation}
At $L=16$ $\mu_l^{eff}$ was found to be 0.99 and increased to the large $l$ 
expected value
1.18 when  $L\sim200$. In other words, $\mu_l^{eff}$ seems to 
interpolate beteen the two limits as expected on the basis of our correction
to scaling.

Middleton himself attributed this gradual evolution of $\mu_l^{eff}$ with $L$ 
to finite size 
corrections which can arise when studying quantities which have
contributions from droplets of all sizes up to the system size $L$. 
The domain wall energy exponent $\theta$ is free of this problem as by 
definition domain walls only exist on the scale $L$. He showed that
even if $E(l)$ varies just as $l^{\theta}$ for all $l$,
then the contributions of droplets
on all length scales up to $L$ means that $\mu_l^{eff}$ approaches its 
asymptotic value only as $1/L^{d-\mu_l}$. These scale averaging corrections
decrease in 2D with the exponent $d-\mu_l\approx0.82$ and are therefore
unlikely to be important when $L\sim100$. In three dimensions 
$d-\mu_l\approx1.3$. As in 3D only modest values of $L$ can be studied  
the scale averaging corrections may there be a significant effect.

The effect of the correction to scaling in the formula for the droplet
energy will have relevance beyond the confines of numerical simulations. The
droplet model provides scaling laws for time-dependent quantities such as 
dynamical susceptibilities in terms of a length scale $L(t)$ which increases 
in a logarithmic manner with time $t$ due to thermal activation of droplets
on the length scale $L(t)$. In practice the length scale which can be 
explored in real experiments is rather limited (typically more than 10 lattice
spacings but probably less than 200 \cite{JYN}). The fact that this 
length scale is not
very large means that corrections to asymptotic formula will be important 
and that will include the correction to the scaling energy 
proposed in this paper.

To summarize: there exists a simple correction to the usual scaling 
formula for the energy of a droplet which is consistent with the TNT 
summary of the numerical data on spin glasses. It would be useful 
to have direct studies of the two correction to scaling exponents
exponents $\omega$ and $\omega^{'}$. 
In 3D and 4D, information on the value of $l$ at the minimum of $E(l)$,
$l_0$, is needed.
As perhaps $l_0$ might have been already seen in the 3D
work of Ref. \cite{LQ}, a precise determination could be possible with a
little extra effort.
In 2D an accurate determination of both exponents 
might be achievable with the techniques used in  Refs. \cite{HY} and \cite{M}.

\begin{acknowledgments}
I would like to thank Alan Bray for many discussions.
\end{acknowledgments}


\begin{thebibliography}{99}
\bibitem{McM}
W. L. McMillan, J. Phys. C \textbf{17}, 3179 (1984).
\bibitem{BM}
A. J. Bray and M. A. Moore, in \textit{ Glassy Dynamics and Optimization},
edited by J. L. van Hemmen and I. Morgenstern, (Springer, Berlin, 1986).
\bibitem{FH}
D. S. Fisher and D. A. Huse, Phys. Rev. Lett. \textbf{56}, 1601 (1986);
Phys. Rev. B \textbf{ 38}, 386 (1988).
\bibitem{Parisi}
G. Parisi, Phys. Rev. Lett. \textbf{43}, 1754 (1979); J. Phys. A \textbf{13},
1101 (1980); Phys. Rev. Lett. \textbf{50}, 1946 (1983).
\bibitem{sgb}
M. M\'{e}zard, G. Parisi and M. A. Virasoro, \textit{Spin Glass Theory and
Beyond} (World Scientific, Singapore, 1987).
\bibitem{KM}
F. Krzakala and O. C. Martin, Phys. Rev. Lett. \textbf{85}, 3013 (2000).
\bibitem{PY}
M. Palassini and A. P. Young, Phys. Rev. Lett. \textbf{85}, 3017 (2000).
\bibitem{Marinari}
E. Marinari, G. Parisi, F. Ricci-Tersenghi, J. Ruiz-Lorenzo and
F. Zuliani, J. Stat. Phys. \textbf{98}, 973 (2000).
\bibitem{3D}
A. J. Bray and M. A. Moore, J. Phys. C \textbf{17}, L463 (1984);
W. L. McMillan, Phys. Rev. B \textbf{30}, 476 (1984); A. K. Hartmann, Phys.
Rev. E \textbf{59}, 84 (1999).
\bibitem{4D}
A. K. Hartmann, Phys. Rev. E \textbf{60}, 5135 (1999); K. Hukushima,
Phys. Rev. E \textbf{60}, 3606 (1999).
\bibitem{Drossel}
M. A. Moore, H. Bokil, and B. Drossel, Phys. Rev. Lett. \textbf{81}, 4252
(1998).
\bibitem{Drossel2}
B. Drossel, H. Bokil, M. A. Moore and A. J. Bray, Eur. Phys. J. B \textbf{13},
369 (2000).
\bibitem{MP}
E. Marinari and G. Parisi, Phys. Rev. Lett. \textbf{86}, 3887 (2001). 
\bibitem{Carter}
A. C. Carter, A. J. Bray and M. A. Moore, Phys. Rev. Lett. \textbf{88}, 077201
(2002).
\bibitem{HKM}
J. Houdayer, F. Krzakala, and O. C. Martin, Eur. Phys. J. B \textbf{18}, 467
(2000).
\bibitem{LQ}
J. Lamarcq, J.-P. Bouchaud, O. C. Martin and M. M\'{e}zard, EuroPhysics Lett.
in press. cond-mat/0107544.
\bibitem{HY}
A. K. Hartmann and A. P. Young, Phys. Rev. B \textbf{64}, 180404 (2001).
\bibitem{KA}
A very extensive review of the 2D spin glass problem has been given by
N. Kawashima and T. Aoki, J. Phys. Soc. Jpn., \textbf{69}, Suppl. A, 169
(2000).
\bibitem{M}
A. A. Middleton Phys. Rev. B \textbf{63}, 060202 (2001).
\bibitem{JYN}
P. E. J\"{o}nsson, H. Yoshino and P. Nordblad, cond-mat /0203444. 

\end{thebibliography}
\end{document}